\documentclass[twocolumn,showpacs,preprintnumbers, 
               superscriptaddress,nofootinbib]{revtex4}
\usepackage[dvipdfmx]{graphicx}
\usepackage{amsmath,amssymb,bm}

\begin{document}

\title{
Thermal blurring of event-by-event fluctuations provoked by rapidity conversion 
}

\author{Yutaro Ohnishi}
\email{yonishi@kern.phys.sci.osaka-u.ac.jp}
\affiliation{
Department of Physics, Osaka University, Toyonaka, Osaka 560-0043, Japan}

\author{Masakiyo Kitazawa}
\email{kitazawa@phys.sci.osaka-u.ac.jp}
\affiliation{
Department of Physics, Osaka University, Toyonaka, Osaka 560-0043, Japan}
\affiliation{
J-PARC Branch, KEK Theory Center,
Institute of Particle and Nuclear Studies, KEK,
203-1, Shirakata, Tokai, Ibaraki, 319-1106, Japan }

\author{Masayuki Asakawa}
\email{yuki@phys.sci.osaka-u.ac.jp}
\affiliation{
Department of Physics, Osaka University, Toyonaka, Osaka 560-0043, Japan}

\begin{abstract}

We study the effect of thermal blurring caused by the 
use of (momentum-space) rapidity as a proxy of 
coordinate-space rapidity in experimental measurements of conserved charge
fluctuations in relativistic heavy ion collisions.
In theoretical studies assuming statistical mechanics, 
calculated fluctuations are those in a spatial volume.
Experiments, on the other hand, can measure fluctuations 
only in a momentum-space in the final state.
In a standard argument to compare experimental results for 
a momentum space with theoretical studies for a coordinate space,
rapidities of particles are implicitly regarded as equivalent to 
their coordinate-space rapidity. 
We show that the relation of two fluctuations
is significantly altered by the existence of the thermal motion,
i.e. thermal blurring.
We discuss that the thermal blurring can be regarded as a
part of the diffusion process, and the effect can be understood
by studying the rapidity window dependences of fluctuations.
Centrality dependence of the thermal blurring effect is also discussed.

\end{abstract}

\date{\today}
\preprint{J-PARC-TH-0062}

\maketitle

\section{Introduction}

In relativistic heavy ion collisions, bulk fluctuations of 
conserved charges observed by event-by-event analyses are 
among unique hadronic observables which carry information on 
the thermal property of the medium in the early stage 
\cite{Stephanov:1998dy,Asakawa:2000wh,Jeon:2000wg,Koch:2008ia};
see a recent review Ref.~\cite{Asakawa:2015ybt}.
In particular, the non-Gaussianity of fluctuations characterized by
higher order cumulants has acquired much attention recently
\cite{Ejiri:2005wq,Stephanov:2008qz,Asakawa:2009aj,Friman:2011pf,
Kitazawa:2013bta,Herold:2016uvv}.
Active measurements of fluctuations have been performed 
at Relativistic Heavy Ion Collider (RHIC) and the Large Hadron
Collider (LHC) \cite{STAR,ALICE,Adamczyk:2013dal,Adamczyk:2014fia,
PHENIX,Luo:2015ewa}.
Measurements will also be carried out in future experiments,
such as the beam-energy scan II (BES-II) program at RHIC \cite{BES-II}
and future facilities, FAIR \cite{FAIR}, NICA \cite{NICA}, 
and J-PARC \cite{J-PARC-HI}.
Conserved-charge fluctuations can 
also be investigated in numerical experiments on the lattice
\cite{Ding:2015ona,Borsanyi:2015axp}.
The comparison between the real and virtual experiments by means of
fluctuations will deepen our knowledge on statistical and
dynamical aspects of relativistic heavy ion collisions.

In the comparison of fluctuations measured by
event-by-event analyses with those obtained by theoretical analyses, 
however, there is a difficulty associated with the phase space 
in which the fluctuation are defined 
\cite{Asakawa:2000wh,Jeon:2000wg,Koch:2008ia}.
On the theoretical side including lattice QCD numerical simulations, 
the cumulants characterizing fluctuations 
are usually calculated on the basis of statistical mechanics
\cite{Stephanov:1998dy,Asakawa:2000wh,Jeon:2000wg,Ejiri:2005wq,
Stephanov:2008qz,Asakawa:2009aj,Friman:2011pf,Ding:2015ona}.
The cumulants calculated in this formalism correspond to 
those in a finite {\it spatial volume} in equilibrium;
the phase space is defined in {\it coordinate space} 
after integrating out the momentum \cite{Asakawa:2015ybt}.
On the other hand, in heavy ion collisions experimental 
detectors cannot observe the position of particles 
in the medium.
Instead, they can only measure the momentum of particles 
in the final state.
Therefore, the phase space defining fluctuations inevitably has 
to be chosen in {\it momentum space}.

The fluctuations in a momentum phase space observed experimentally 
are usually regarded as a proxy of the one in a coordinate space
as follows \cite{Asakawa:2000wh,Jeon:2000wg}.
First, assuming the Bjorken space-time evolution 
the (momentum-space) rapidity\footnote{
Pseudorapidity is often experimentally measured insted of
rapidity because of the relative easiness of the measurement.
In this paper, however, we consider rapidity because
theoretically it has a preferable feature under
Lorentz boost.
}
$y$ of a {\it fluid element} is 
equivalent to the coordinate-space rapidity $Y=\tanh^{-1}(z/t)$ 
of the fluid element because of boost invariance, where $t$ and $z$ 
are time and the longitudinal coordinate, respectively.
Second, by assuming that the rapidities of {\it individual particles} 
in the fluid element is equivalent to the rapidity of the fluid element,
rapidities of particles are identical with $Y$.
Then, by measuring fluctuations in a rapidity window $\Delta y$ 
after integrating out the transverse momentum,
the phase space is regarded as the one in the coordinate space 
in a coordinate-space rapidity window $\Delta Y = \Delta y$,
where transverse coordinates, $x$ and $y$, are integrated out.

This argument, however, relies on two nontrivial assumptions;
(1) validity of the Bjorken picture and (2) that the relative velocities 
of individual particles against the fluid element are negligible.
Though the former may be justified for sufficiently high energy
collisions, the latter can be invalidated by thermal motion irrespective
of collision energy.
Because of the thermal motion, 
the correspondence between the two rapidities $y$ and $Y$ 
for individual particles becomes at most an approximate one.
The measurement of fluctuation in $\Delta y$ thus receives a blurring 
effect when the results are to be interpreted as fluctuations 
in $\Delta Y$.
In this study, we call this effect as thermal blurring,
and investigate its effect on fluctuation observables quantitatively.
We note that the existence of the thermal blurring effect 
has been pointed out in earlier studies \cite{Asakawa:2000wh,Jeon:2000wg,
Shuryak:2000pd,Koch:2008ia}.
The same problem is recently investigated in a slightly different 
context in Ref.~\cite{Ling:2015yau}.
The purpose of the present study is to investigate this effect
on cumulants quantitatively.
We discuss the centrality dependence of the thermal blurring effect, 
and extend the argument to non-Gaussian fluctuations.
The main results of this paper are presented 
in Ref.~\cite{Asakawa:QM2015}.

In this study, we estimate the thermal blurring effect by 
assuming that individual particles are emitted from the medium 
at kinetic freezeout.
The thermal motion of individual particles at kinetic freezeout 
is deduced from a simple blastwave model for particle yields
in $p_{\rm T}$ space.
We show that the thermal blurring effect becomes more prominent 
as the rapidity window $\Delta y$ becomes narrower, and 
at the maximal coverage of the rapidity window of STAR detector
the observed fluctuations are significantly modified owing to this effect.

Because we consider the thermal blurring at kinetic freezeout, 
our argument relates fluctuations observed experimentally 
to those in $\Delta Y$ at kinetic freezeout.
When one wants to compare the experimental results 
with thermal fluctuations generated in much earlier stage,
one has to take account of the time evolution of fluctuations 
before kinetic freezeout \cite{
Shuryak:2000pd,Kitazawa:2013bta,Kitazawa:2015ira}.
The time evolution is basically the diffusion process toward
the equilibrium.
In this paper we discuss that the thermal blurring can be 
regarded as a part of the diffusion process.
One thus can use the mathematical results in Refs.~\cite{
Shuryak:2000pd,Kitazawa:2013bta,Kitazawa:2015ira} directly
to understand the thermal blurring effects.
We argue that the modification of fluctuations 
due to thermal blurring and diffusion can be experimentally 
understood by studying the rapidity window $\Delta y$ dependences 
of the cumulants as discussed for the case of diffusion in 
Refs.~\cite{Kitazawa:2013bta,Sakaida:2014pya,Kitazawa:2015ira}.
The centrality dependence of net-electric charge fluctuation 
observed by ALICE collaboration \cite{ALICE}
is also discussed on the basis of this picture.

Throughout this paper, we assume the Bjorken space-time evolution.
At lower energy collisions, this picture does not hold 
and our discussion would be significantly modified.
We, however, do not consider such effects until 
Sec.~\ref{sec:summary}.

This paper is organized as follows.
In the next section we study thermal distribution of particles
in rapidity space using a simple blastwave model.
In Sec.~\ref{sec:<Q^n>}, we then study the thermal blurring
effects on cumulants. 
The formula of the cumulants are derived with two
different methods in Secs.~\ref{sec:binomial} and \ref{sec:multinomial}.
Numerical results are then shown in Secs.~\ref{sec:num} and
\ref{sec:centrality}.
In Sec.~\ref{sec:D+B}, we then consider the effect of diffusion 
in the hadronic stage and show that the diffusion and blurring 
can be regarded as parts of a single diffusion process on the same footing.
Section~\ref{sec:summary} is devoted to discussions and a short summary.

\section{Thermal distribution in rapidity space}
\label{sec:n(y)}

In this section, we first discuss the magnitude of 
thermal blurring by studying the thermal distribution of 
particles in rapidity space at kinetic freezeout on the basis of 
a blastwave model.

In the Bjorken space-time evolution, 
the distribution of particle density in $y$ space 
at coordinate-space rapidity $Y$, $n_Y(y)$ is related to 
the distribution at mid-rapidity $n(y)$ as 
\begin{align}
n_Y(y) = n(y-Y) ,
\end{align}
because of boost invariance.
In what follows, we thus concentrate on $n(y)$.

The invariant momentum spectrum of particles
crossing a surface element $d\Sigma_\mu$ is given by 
the Cooper-Frye formula \cite{Cooper:1974mv},
\begin{align}
E\frac{d N}{d^3 \bm{p} } = d\Sigma\cdot p f(p\cdot u ) ,
\label{eq:Cooper-Frye}
\end{align}
where $f(E)$ is the single-particle distribution 
in the rest frame and $u_\mu$ denotes the flow velocity.
We assume the Boltzmann distribution for $f(E)$
\begin{align}
f(E) \sim \exp\bigg[ -\frac{E-\mu}T \bigg],
\label{eq:Boltzmann}
\end{align}
with the temperature $T$, chemical potential $\mu$, and 
$E=\sqrt{m^2+\bm{p}^2}$ with $m$ denoting 
the mass of particles.
The effect of quantum statistics on Eq.~(\ref{eq:Boltzmann})
is well suppressed for $T\ll m-\mu$.
At kinetic freezeout point with the temperature $T_{\rm kin}$, 
the effect of quantum statistics is negligible for all particles 
except for pions, on which the effect is at most about $10\%$.

In order to calculate the particle distribution $n(y)$ with 
Eq.~(\ref{eq:Cooper-Frye}), 
we employ the following simplified blastwave model:
We assume that the freezeout with temperature $T_{\rm kin}$ 
and chemical potential $\mu_{\rm kin}$ takes place 
at a fixed proper time $\tau_{\rm kin}$ with a constant transverse 
velocity $\beta$ \footnote{
In this model, therefore, possible dependence of $\tau_{\rm kin}$
on the position in transverse plane is neglected.
The possible azimuthal angle dependence of $\beta$ for peripheral 
collisions is not taken into account in this model, either.}.
The flow vector at mid-rapidity at
$(t,x,y,z) = (t_{\rm kin},x,0,0)$
is given by $u_\mu = ( \gamma , \beta\gamma , 0 , 0 )$
with $\gamma=(1-\beta^2)^{-1/2}$ 
while the surface vector $d\Sigma_\mu$ is proportional to 
$(1,0,0,0)$.
Substituting them into Eq.~(\ref{eq:Cooper-Frye}),
the momentum distribution of the emitted particles from
the freezeout surface at this point is given by
\begin{align}
\frac{d N}{d^3 \bm{p} } 
\sim \exp [ - p\cdot u /T ] 
= \exp [ - \gamma ( E + \beta p_x ) /T ]~,
\end{align}
where in the first proportionality we have used the fact that the 
$\mu$ dependence through $\exp(\mu/T)$ can be factored out.
The particle spectrum emitted from 
freezeout surface per unit rapidity $y$ per unit
transverse momentum $\bm{p}_{\rm T}$ is given by 
\begin{align}
\tilde{n}(\bm{p}_{\rm T},y) = \frac{d N}{d^3 \bm{p} } \frac{dp_z}{dy}.
\end{align}
Using $d p_z = E dy $ and 
by integrating out the transverse momentum,
we obtain the particle distribution per unit rapidity as
\begin{align}
n(y) \sim 
\int dp_x dp_y E e^{-\gamma( E+\beta p_x )/T } ,
\label{eq:n(y)}
\end{align}
where we have used the rotational invariance with regard to
$z$ axis, i.e. the longitudinal axis.
The proportionality coefficient of $n(y)$ is determined so as to 
satisfy $\int dy n(y)=1$.
We note that Eq.~(\ref{eq:n(y)}) does not depend on $\mu$.
It is easily shown that $n(y)$ depends on 
$m$ and $T$ only through the combination $w=m/T$.

\begin{figure}
\begin{center}
\includegraphics[width=.49\textwidth]{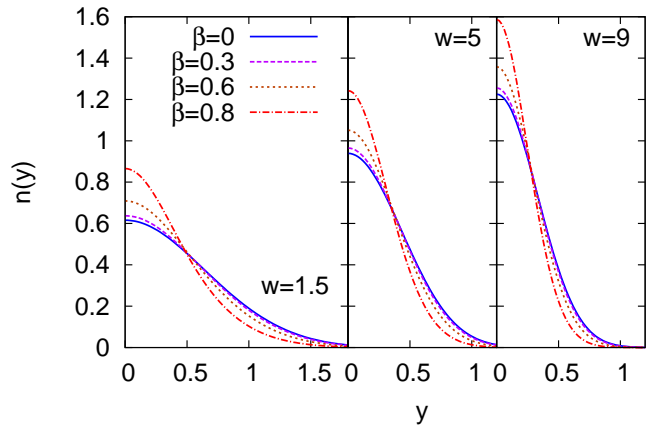}
\caption{
Particle density per unit rapidity $n(y)$
for several values of $w=m/T$ and the transverse velocity $\beta$.
}
\label{fig:y-dist}
\end{center}
\end{figure}

In Fig.~\ref{fig:y-dist} the distribution $n(y)$ is plotted 
for several values of $w=m/T$ and $\beta$.
The figure shows that the distribution becomes narrower as 
$w$ becomes larger. This dependence comes from the 
suppression of thermal motion at large $w$.
The figure also shows that the distribution becomes narrower 
for large $\beta$, which is a consequence of Lorentz
effect; with the boost of a thermal system, 
the distribution is squeezed toward the direction of the boost.

\begin{figure}
\begin{center}
\includegraphics[width=.49\textwidth]{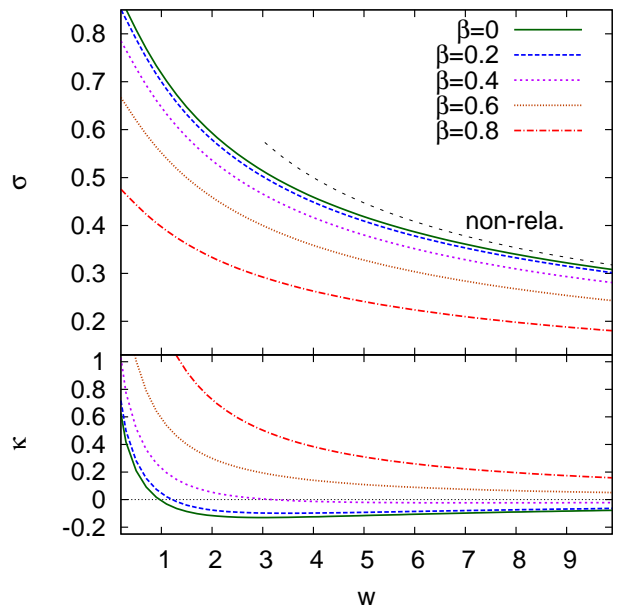}
\caption{
Width $\sigma$ and kurtosis $\kappa$ of $n(y)$ 
as a function of $w=m/T$ for several values of 
transverse velocity $\beta$.
}
\label{fig:width_w}
\end{center}
\end{figure}

\begin{figure}
\begin{center}
\includegraphics[width=.49\textwidth]{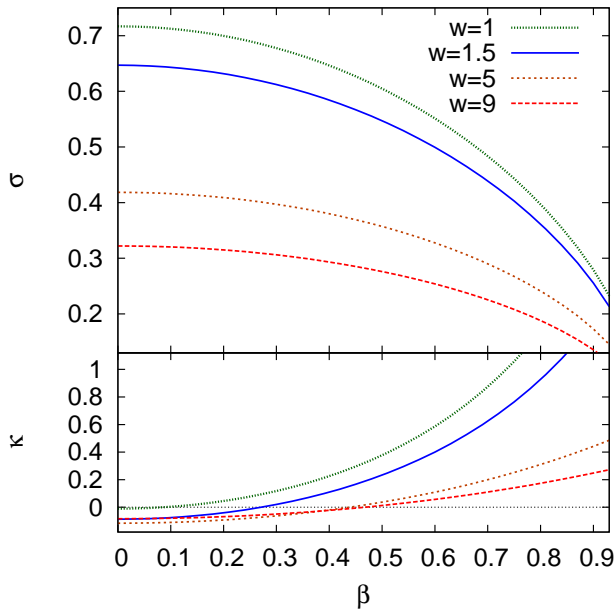}
\caption{
Width $\sigma$ and kurtosis $\kappa$ of $n(y)$ 
as a function of transverse velocity $\beta$ 
for several values of $w=m/T$.
}
\label{fig:width_b}
\end{center}
\end{figure}

In order to characterize the thermal distribution more quantitatively, 
we plot the width $\sigma$ of $n(y)$ defined by
\begin{align}
\sigma^2 = \int dy y^2 n(y) ,
\label{eq:sigma}
\end{align}
as functions of $w$ and $\beta$ in the upper panels of
Figs.~\ref{fig:width_w} and \ref{fig:width_b},
respectively.
The blastwave fits for the $p_{\rm T}$ spectra 
at the LHC and top-RHIC energies for the most 
central collisions show that the freezeout parameters are
$T_{\rm kin}\simeq 100$ MeV and $\beta=0.6-0.7$ \cite{Abelev:2013vea}.
With $T=T_{\rm kin}$, we thus have $w=m/T\simeq 1.5$ and $9$
for pions and nucleons, respectively.
Figures~\ref{fig:width_w} and \ref{fig:width_b} show that 
the width of $n(y)$ for pions is $\sigma\simeq0.5$ with the 
blastwave parameters.
This value is almost half the maximal rapidity
window $\Delta y=1.0$ at STAR \cite{Adamczyk:2013dal}.
Because the electric charge is dominantly carried by pions,
this result suggests that the measurement of electric charge
fluctuations with $\Delta y=1.0$ \cite{Adamczyk:2014fia} is 
strongly affected by thermal blurring.
For nucleons, we have $\sigma\simeq0.25$ with the same 
freezeout parameters.
The measurement of the baryon number cumulants \cite{KA} thus
is less affected by thermal blurring than the electric charge,
although the magnitude of $\sigma$ in this case is not much suppressed
compared to $\Delta y=1.0$, either.
In the next section, we analyze the thermal blurring effect
more quantitatively by studying the cumulants directly.

Next, let us consider the deviation of 
$n(y)$ from Gauss distribution.
Typical parameters to represent the deviation are
the skewness $S$ and kurtosis $\kappa$ defined by \cite{Asakawa:2015ybt}
\begin{align}
S &= \frac1{\sigma^3} \int dy y^3 n(y) ,
\\
\kappa &= \frac1{\sigma^4} \int dy y^4 n(y) - 3 .
\end{align}
Because $S$ and $\kappa$ vanish for the Gauss distribution, 
their nonzero values characterize non-Gaussianity\footnote{
Here, we emphasize that $S$ and $\kappa$ defined here are 
the skewness and kurtosis {\it of} $n(y)$, respectively, and thus are different
from those of event-by-event fluctuations of a conserved charge.}. 
Since $n(y)$ is an even function, $S$ always vanishes.
In the lower panels of Figs.~\ref{fig:width_w} and \ref{fig:width_b}, 
the kurtosis of $n(y)$ is plotted for various parameters.
Because the Maxwell-Boltzmann distribution in non-relativistic
gas is given by Gaussian, non-Gaussianity of $n(y)$ comes from 
relativistic effects. In fact, 
the figures show that the magnitude of $\kappa$ becomes 
large for small $w$ and large $\beta$, at which the 
relativistic effects become more prominent.
For the parameters relevant to pions and nucleons at 
kinetic freezeout, however, we have $|\kappa|<0.5$, which indicates 
that the deviation from the Gauss distribution is not large.
In Sec.~\ref{sec:num} we will show that the effect of 
non-Gaussianity of $n(y)$ on cumulants is indeed
well suppressed.

In the non-relativistic limit $w\to\infty$ and $\beta\to0$,
the distribution $n(y)$ approaches a Gauss distribution
with the width $\sigma=1/w$, which is shown by 
the dotted line in Fig.~\ref{fig:width_w}.

\section{Blurring effect on cumulants}
\label{sec:<Q^n>}

Next, we investigate the effects of thermal blurring 
on cumulants of a particle number $Q_{\Delta y}$ in 
a rapidity window $\Delta y$.
To this end, in this section we first develop
the formulation for the cumulants of $Q_{\Delta y}$
using two different methods, which give the same result.
In Sec.~\ref{sec:binomial}, we first derive the result by only using 
the general properties of the cumulants and the binomial distribution
function. We then obtain the same result in Sec.~\ref{sec:multinomial} 
starting from a discretized formalism.

In this study, 
we investigate the thermal blurring effect focusing on 
the case that the density in $Y$ space before thermal blurring 
is given by $\rho_{\rm Y}(Y)$ and does not have event-by-event fluctuation.
The density $\rho_{\rm y}(y)$ in $y$ space after thermal 
blurring has event-by-event fluctuations even in this case, 
and accordingly the cumulants $\langle (Q_{\Delta y})^n \rangle_{\rm c}$ 
for $n\ge2$ have nonzero values.
Because $\langle (Q_{\Delta y})^n \rangle_{\rm c}$ for $n\ge2$
vanish without thermal blurring in this case, their nonzero values 
can be used as a measure of the magnitude of this effect.
As we will discuss in Sec.~\ref{sec:DME}, 
the fluctuations of $\rho_{\rm Y}(Y)$ can straightforwardly be 
incorporated in this analysis following the treatment in 
Refs.~\cite{Kitazawa:2013bta,Kitazawa:2015ira}.

\subsection{Simple derivation}
\label{sec:binomial}

\begin{figure}
\begin{center}
\includegraphics[width=.4\textwidth]{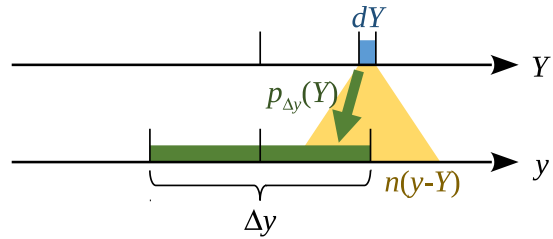}
\caption{
Illustration of the rapidity window $\Delta y$ and 
the probability $p_{\Delta y}(Y)$ in Eq.~(\ref{eq:p_DyY}).
}
\label{fig:N_binomial}
\end{center}
\end{figure}

In order to describe the cumulants of $Q_{\Delta y}$, 
we first consider particles in an infinitesimal range 
$dY$ in $Y$ space before thermal blurring.
The number of particles in $dY$ is 
\begin{align}
N_{dY}(Y) = \rho_{\rm Y}(Y)dY.
\end{align}
After thermal blurring, a particle in $dY$ 
is found in the rapidity interval $-\Delta y/2\le y \le \Delta y/2$
with probability
\begin{align}
p_{\Delta y}(Y) 
= \int_{-\Delta y/2}^{\Delta y/2} dy n(y-Y) .
\label{eq:p_DyY}
\end{align}
See Fig.~\ref{fig:N_binomial}.
Because of the nature of thermal blurring,
this probability is to be regarded independent
for individual particles.
Therefore, the distribution of the particle number 
found in the rapidity interval $\Delta y$ 
\begin{align}
q_{dY\to\Delta y} &\mbox{: Number of particles which exist in $dY$}
\nonumber \\
&~~\mbox{{\it and} are found in $\Delta y$,}
\end{align}
obeys the
binomial distribution function,
\begin{align}
B_{p,N}(m) = {_N C}_m p^m (1-p)^{N-m},
\label{eq:binomial}
\end{align}
with $p=p_{\Delta y}(Y)$ and $N=N_{dY}(Y)$.
Using the fact that the $n$th order cumulant of 
Eq.~(\ref{eq:binomial}) is given by \cite{Asakawa:2015ybt} 
\begin{align}
\langle m^n \rangle_{\rm c} = \xi_n(p) N,
\label{eq:xi}
\end{align}
we find that the cumulants of $q_{dY\to\Delta y}$ are given by 
\begin{align}
\big\langle ( q_{dY\to\Delta y} )^n \big\rangle_{\rm c}
&= \xi_n(p_{\Delta y}(Y)) N_{dY}(Y)
\nonumber \\
&= \xi_n(p_{\Delta y}(Y)) \rho_{\rm Y}(Y)dY.
\label{eq:<q^l>}
\end{align}
The explicit forms of $\xi_n(p)$ up to the fourth order are 
\begin{align}
\xi_1(p) &=p, ~
\xi_2(p)=p(1-p), ~
\xi_3(p)=p(1-p)(1-2p), 
\nonumber \\
\xi_4(p) &= p(1-p)(1-6p+6p^2).
\end{align}
For the fifth and sixth orders, see Ref.~\cite{Kitazawa:2016awu}.

The total number of particles $Q_{\Delta y}$ in $\Delta y$ 
is obtained by summing up $q_{dY\to\Delta y}$ for all infinitesimal 
ranges $dY$ as 
\begin{align}
Q_{\Delta y} = \sum_{\{dY\}} q_{dY\to\Delta y}.
\end{align}
To calculate the cumulants of $Q_{\Delta y}$, 
we note that the cumulants of the sum of uncorrelated 
stochastic variables are simply given by the sum of the cumulants 
\cite{Asakawa:2015ybt}.
Because $q_{dY\to\Delta y}$ should be uncorrelated 
for different $dY$ bins,
the cumulants of $Q_{\Delta y}$ are obtained as
\begin{align}
\langle (Q_{\Delta y})^n \rangle_{\rm c}
&= \sum_{\{dY\}} \langle (q_{dY\to\Delta y})^n \rangle_{\rm c}
\nonumber \\
&= \int dY \xi_n(p_{\Delta y}(Y)) \rho_{\rm Y}(Y),
\label{eq:<Q^n>}
\end{align}
where in the last equality we used Eq.~(\ref{eq:<q^l>}), 
and replaced the sum with an integral.
When the density $\rho_{\rm Y}(Y)$ is uniform,
$\rho_{\rm Y}(Y) = \rho_0$, we have 
\begin{align}
\langle (Q_{\Delta y})^n \rangle_{\rm c}
= \rho_0 \int dY \xi_n(p_{\Delta y}(Y)).
\label{eq:<Q^n>uniform}
\end{align}

Next, let us see the behavior of Eq.~(\ref{eq:<Q^n>})
in the $\Delta y \to 0$ limit.
In this limit, the probability $p_{\Delta y}(Y)$ should be suppressed
proportionally to $\Delta y$ irrespective of the value of $Y$.
Because $\xi_n(p)$ in Eq.~(\ref{eq:xi}) satisfy
$\xi_n(p)\to p$ for $p\to0$ \cite{Asakawa:2015ybt}, 
we have
\begin{align}
\xi_n(p_{\Delta y}(Y)) \to p_{\Delta y}(Y) 
\quad \mbox{for $\Delta y \to 0$},
\end{align}
and the cumulants of $Q_{\Delta y}$ converge to a common value
\begin{align}
\langle (Q_{\Delta y})^n \rangle_{\rm c}
= \int dY p_{\Delta y}(Y) \rho_{\rm Y}(Y) = \langle Q_{\Delta y} \rangle,
\label{eq:Dy->0}
\end{align}
for any $n\ge1$.
This result shows that $Q_{\Delta y}$ in this limit obeys
a Poisson distribution \cite{Asakawa:2015ybt}.

For small but finite $\Delta y$, the probability $p_{\Delta y}(Y)$ 
may be expanded by a power series of $\Delta y$
starting from the first order.
By substituting this expansion into Eq.~(\ref{eq:<Q^n>})
and using Eq.~(\ref{eq:xi}), 
one finds that the $\Delta y$ dependence of 
$\langle (Q_{\Delta y})^n \rangle_{\rm c}$ is also expanded by 
a power series of $\Delta y$.
The $n$th order term in this expansion for $n\ge2$
generally takes a nonzero value.

\subsection{Derivation based on discretized formalism}
\label{sec:multinomial}

\begin{figure}
\begin{center}
\includegraphics[width=.4\textwidth]{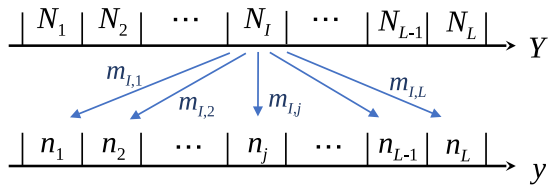}
\caption{
Discrete system discussed in Sec.~\ref{sec:multinomial}.
}
\label{fig:N_discrete}
\end{center}
\end{figure}

Next, we derive Eq.~(\ref{eq:<Q^n>}) again but in a different way.
In this subsection, we start from discretized  
coordinate spaces and take the continuum limit at the end.

We divide the coordinates $Y$ and $y$ into discrete cells 
with equal lengths $\delta Y$ and $\delta y$, respectively,
as illustrated in Fig.~\ref{fig:N_discrete}.
For simplicity, we further assume $\delta Y = \delta y$,
though this is not essential in the following argument.
For the moment, we assume that the total number of the cells 
$L$ is finite in each space, although the final result does not depend 
on the finiteness of the number of cells.
The distribution of particles in $Y$ space is represented 
by the numbers of particles $N_I$ in individual 
cells labeled by $I=1,2,\cdots,L$.
The distribution after thermal blurring in $y$ space 
is also represented by the number of particles $n_j$ 
in cells labeled by $j=1,2,\cdots,L$.

As in the previous subsection, 
we consider the thermal blurring starting from a fixed
distribution $\bm{N}=(N_1,\cdots,N_L)$ in $Y$ space.
Thermal blurring gives rise to fluctuation of 
the distribution $\bm{n}=(n_1,\cdots,n_L)$ in $y$ space.
We denote the probability distribution function of 
$\bm{n}$ by $P_y(\bm{n})$.

Owing to thermal blurring, 
a particle in a cell, say in the $I$th cell, in $Y$ space is
distributed to various cells in $y$ space.
We denote the probability that the particle 
is found in $j$th cell in $y$ space as 
\begin{align}
P_{I\to j} ~&\mbox{: the probability that a particle}
\nonumber \\
&~\mbox{  in the $I$th cell in $Y$ space is found}
\nonumber \\
&~\mbox{  in the $j$th cell in $y$ space}.
\label{eq:P_ij}
\end{align}
Note that $\sum_j P_{I\to j} = 1$ has to be satisfied.
Next, we consider the probability that $N_I$ particles in 
the $I$th cell are distributed in $y$ space with $m_{I,j}$ particles 
in the $j$th cell as shown in Fig.~\ref{fig:N_discrete}.
This probability is given by 
\begin{align}
p_I (\bm{m}_I;N_I) = f_{N_I}(\bm{m}_I ; \bm{P}_I),
\label{eq:p_i}
\end{align}
with $\bm{m}_I=(m_{I,1},\cdots,m_{I,L})$, 
$\bm{P}_I=(P_{I\to 1},\cdots,P_{I\to L})$, and 
the multinomial distribution function
\begin{align}
f_{N_I}(\bm{m}_I ; \bm{P}_I)
= \frac{N_I !}{m_{I,1}! \cdots m_{I,L}!}\prod_j (P_{I\to j})^{m_{I,j}}
\cdot \delta_{N_I,\sum_j m_{I,j}},
\label{eq:f_Ni}
\end{align}
where the Kronecker delta represents 
the conservation of particle number.
The probability $P_y(\bm{n})$ is then given by the product of 
Eq.~(\ref{eq:p_i}) after the sum over $m_{I,j}$ for all cells 
with a constraint $n_j=\sum_I m_{I,j}$ as
\begin{align}
P_y(\bm{n})
= \prod_I \bigg[ \sum_{\bm{m}_I} p_I (\bm{m}_I)  \bigg]
\cdot \prod_j \delta_{n_j,\sum_I m_{I,j}}.
\label{eq:P_y(n)}
\end{align}

To calculate the cumulants of $\bm{n}$,
it is convenient to use generating functions \cite{Asakawa:2015ybt}.
The factorial moment generating function of Eq.~(\ref{eq:P_y(n)})
is calculated to be
\begin{align}
G_{\rm f}(\bm{s}) 
&= \sum_{\bm{n}} \bigg[ \big( \prod_j s_j^{n_j} \big) P_y(\bm{n}) \bigg]
\nonumber \\
&= \prod_I \bigg[ \sum_{\bm{m}_I} p_i (\bm{m}_I) \prod_j s_j^{m_{I,j}} \bigg] 
\nonumber \\
&= \prod_I \big( \sum_j s_j P_{I\to j} \big)^{N_I} ,
\end{align}
with $\bm{s}=(s_1,\cdots,s_L)$. In the last step,
we used Eqs.~(\ref{eq:p_i}) and (\ref{eq:f_Ni}).
The factorial cumulant generating function is then obtained as 
\begin{align}
K_{\rm f}(\bm{s})  = \ln G_{\rm f}(\bm{s}) 
= \sum_I N_I \ln \big( \sum_j s_j P_{I\to j} \big) .
\label{eq:K_f}
\end{align}

To take the continuum limit, $\delta Y \to 0 $ and 
$\delta y\to0$, of Eq.~(\ref{eq:K_f}), we replace
$N_i\to \rho_{\rm Y}(Y) \delta Y$ and $P_{I\to j}=n(y-Y)\delta y$.
The sums over $I$ and $j$ in Eq.~(\ref{eq:K_f}) 
then become integrals and one obtains the generating
functional
\begin{align}
K_{\rm f}[s(y)]
= \int dY \rho_{\rm Y}(Y) \ln \big[ \int dy s(y) n(\eta-Y) \big].
\label{eq:K_f_cont}
\end{align}

The factorial cumulants of $Q_{\Delta y}$ are obtained by applying
the operator
\begin{align}
D_{\Delta y} = \int_{-\Delta y/2}^{\Delta y/2} dy' \frac{\delta}{\delta s(y')} ,
\end{align}
to Eq.~(\ref{eq:K_f_cont}) and taking $s(y)=1$ afterwards as 
\begin{align}
\langle (Q_{\Delta y})^n \rangle_{\rm fc} = (D_{\Delta y})^n K_{\rm f}|_{s(y)=1} ,
\end{align}
where the definition of 
the functional derivative $\delta/\delta s(y)$ is 
understood as the limit of the discretized notation.
The cumulants of $Q_{\Delta y}$ are then obtained by using the 
relation between cumulants and factorial cumulants 
\cite{Kitazawa:2015ira,Asakawa:2015ybt}.
This manipulation leads to Eq.~(\ref{eq:<Q^n>}).
We note that with $s(y)=1$ 
the argument of logarithmic function in Eq.~(\ref{eq:K_f_cont})
becomes unity, $\int dy s(y) n(\eta-Y)=\int dy n(\eta-Y)=1$,
which makes the manipulation apparent.

\subsection{Relation with diffusion master equation and initial fluctuation}
\label{sec:DME}

Here, we note that Eq.~(\ref{eq:<Q^n>}) has the same form
as the results of the cumulants $\langle (Q_{\Delta y})^n \rangle_{\rm c}$ 
obtained in the diffusion master equation (DME) 
\cite{Kitazawa:2013bta,Kitazawa:2015ira}
with fixed initial condition when $n(y)$ is replaced by
a Gauss distribution with the width
$( 2 \int_0^t dt' D(t) )^{1/2}$, where $D(t)$ is 
the time ($t$) dependent diffusion coefficient;
see, Sec.~2.5 in \cite{Kitazawa:2015ira}.
This correspondence is reasonable, because in the DME 
individual particles composing the system behave independently,
and the location of a particle at time $t$ with a fixed initial 
position is distributed by a Gauss distribution owing to their 
random motion. Note that the Gaussianity of this distribution in 
the DME is consistent with the particle diffusion described by 
a diffusion equation \cite{Kitazawa:2015ira,Asakawa:2015ybt}.

The results in Eq.~(\ref{eq:<Q^n>}) are obtained for fixed 
initial density $\rho_{\rm Y}(Y)$ without fluctuation.
In Refs.~\cite{Kitazawa:2013bta,Kitazawa:2015ira}, the 
solutions of the DME are obtained for initial conditions 
including fluctuations; see Sec.~2.7 in Ref.~\cite{Kitazawa:2015ira}
for example.
The derivation in these studies is applicable straightforwardly 
to the present problem, thermal blurring.
When the fluctuation of $\rho_{\rm Y}(Y)$ is taken into account, 
the result Eq.~(\ref{eq:<Q^n>}) 
for the first and second order cumulants is modified as
\begin{align}
\langle Q_{\Delta y} \rangle_{\rm c}
&= \int dY \langle \rho_{\rm Y}(Y)\rangle_0 p_{\Delta y}(Y) ,
\\
\langle (Q_{\Delta y})^2 \rangle_{\rm c}
&= \int dY \langle \rho_{\rm Y}(Y)\rangle_0 
\xi_2( p_{\Delta y}(Y) )
\nonumber \\
&+ \int dY_1dY_2 \langle \delta\rho_{\rm Y}(Y_1) \delta\rho_{\rm Y}(Y_2) \rangle_0
p_{\Delta y}(Y_1) p_{\Delta y}(Y_2), 
\label{eq:Q2fluc}
\end{align}
where $\langle \cdot \rangle_0$ represents the expectation values 
taken for the distribution of $\rho_{\rm Y}(Y)$ with 
$\delta\rho_{\rm Y}(Y) = \rho_{\rm Y}(Y) - \langle \rho_{\rm Y}(Y) \rangle_0$.
In Refs.~\cite{Kitazawa:2013bta,Kitazawa:2015ira}, 
the result is also extended to describe
the net-particle number, i.e. the difference of the particle
and anti-particle numbers.

\section{Numerical Results}
\label{sec:num}

\subsection{Rapidity window dependence}
\label{sec:num-dy}

\begin{figure}
\begin{center}
\includegraphics[width=.49\textwidth]{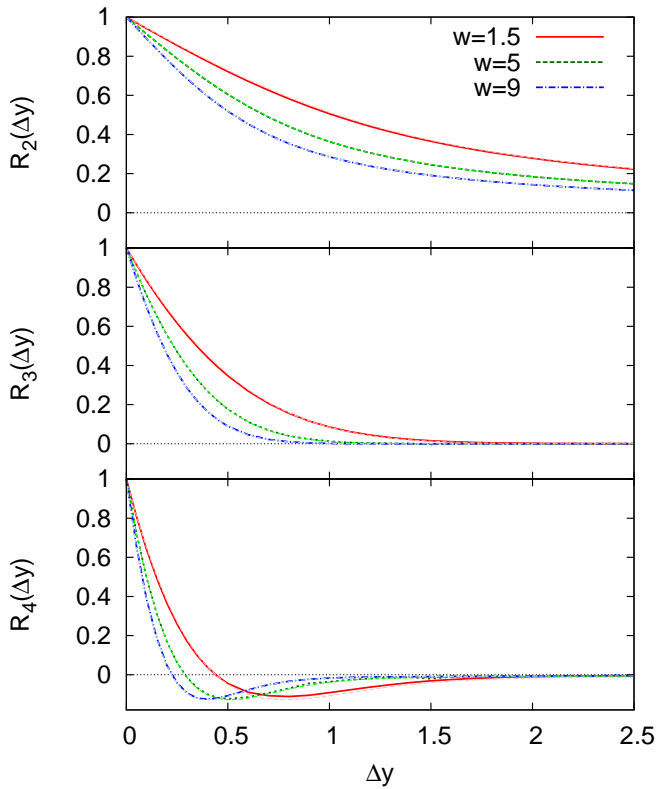}
\caption{
Rapidity window dependences of the cumulants in normalization
$R_n = \langle Q_{\Delta y}^n \rangle_{\rm c}/\langle Q_{\Delta y} \rangle_{\rm c}$
with vanishing 
initial condition for several values of $w=m/T$ with $\beta=0.6$.
}
\label{fig:Dy}
\end{center}
\end{figure}

Next, we see the thermal blurring effect on 
$\langle (Q_{\Delta y})^n \rangle_{\rm c}$ numerically.
In this section, we consider the cumulants for homogeneous 
distribution in $Y$ space, Eq.~(\ref{eq:<Q^n>uniform}).
In Fig.~\ref{fig:Dy}, we show the $\Delta y$ dependences of
the ratio of the cumulants normalized by the Poissonian value
\cite{Kitazawa:2015ira}
\begin{align}
R_n(\Delta y) = \frac{\langle (Q_{\Delta y})^n \rangle_{\rm c}}
{\langle Q_{\Delta y} \rangle_{\rm c}},
\label{eq:R_n}
\end{align}
for several values of $w$ with $\beta=0.6$ \cite{Abelev:2013vea}.
Because $R_n(\Delta y)$ should vanish without thermal blurring,
their nonzero values represent a measure to see
the magnitude of the thermal blurring effect.
The figure shows that $R_n(\Delta y)$ for $n\ge2$ becomes unity
in the limit $\Delta y\to0$.
This result is consistent with Eq.~(\ref{eq:Dy->0}), which states
that the distribution of $Q_{\Delta y}$ becomes Poissonian in this limit.
This limit can be regarded as the case that the information
on the event-by-event fluctuations of $\rho_{\rm Y}(Y)$ is completely 
lost owing to thermal blurring.
On the other hand, $R_n(\Delta y)$ approaches zero for large $\Delta y$.
This result shows that the thermal blurring effect is suppressed
when $\Delta y\gg\sigma$ is satisfied.
We note that the $\Delta y$ dependence of the cumulants in 
Fig.~\ref{fig:Dy} can be compared with the experimental results.
For example, $R_2(\Delta y)$ is related to the D-measure $D$ as 
$R_2(\Delta y)= D/4$ \cite{Koch:2008ia}.

The maximal rapidity window of STAR detector is $\Delta y=1.0$.
The upper panel of Fig.~\ref{fig:Dy} shows that 
the effect of thermal blurring is rather strong for the second order 
cumulant with $\Delta y=1.0$. For $w=1.5$ corresponding to the 
electric charge fluctuation, we have $R_2(\Delta y)\simeq0.5$.
This result shows that the second-order cumulant of
the electric charge fluctuation $\langle N_{\rm Q}^2 \rangle_{\rm c}$ 
observed by STAR \cite{Adamczyk:2014fia}
receives about $50\%$ modification due to thermal blurring.
Because $\langle N_{\rm Q}^2 \rangle_{\rm c}$ is modified, the ratio of 
the cumulants $\kappa\sigma^2=\langle N_{\rm Q}^4 \rangle_{\rm c}/
\langle N_{\rm Q}^2 \rangle_{\rm c}$ is also modified.
The blurring effect is smaller at the maximal rapidity window of 
the ALICE detector $\Delta y=1.6$ \cite{ALICE} or 
for $w=9$ corresponding to net-baryon number cumulants.
Even for these cases, however, $R_2(\Delta y)$ is larger than $0.25$,
which indicates that the thermal blurring effect is 
not well suppressed.
The middle and lower panels of Fig.~\ref{fig:Dy} show
the results for the third and fourth order cumulants, 
$R_3(\Delta y)$ and $R_4(\Delta y)$.
These panels suggest that the thermal blurring effect is 
more suppressed for higher order, but is not negligible 
for $\Delta y=1.0$.
In particular, the sign of the fourth order cumulant 
can become negative owing to this effect.

From the results in Fig.~\ref{fig:Dy}, it is interesting 
to analyze the $\Delta y$ dependences of the cumulants experimentally 
and compare with these results.
In particular, the simultaneous analysis of the second, third, and 
fourth order cumulants for electric charge and baryon number 
cumulants would enable us to confirm the validity of the picture 
on thermal blurring, and to investigate its magnitude.
We also note that the wider rapidity coverage is desirable
for the analysis of $\Delta y$ dependences.
The extension of STAR detector to cover wider $\Delta y$ 
\cite{BES-II} thus will be quite effective for these analyses.

In order to see the effect of non-Gaussianity 
of $n(y)$ on our results, we calculate 
$\langle (Q_{\Delta y})^n \rangle_{\rm c}$ by replacing 
$n(y)$ with a Gauss function with the width in Eq.~(\ref{eq:sigma}).
The results are shown by the lines with light colors 
in Fig.~\ref{fig:Dy}.
The difference of these results from those with $n(y)$, however, 
is small and almost invisible in the figure except for 
$\langle (Q_{\Delta y})^4 \rangle_{\rm c}$ with $w=1.5$
having a small deviation.
This result shows that one can safely replace $n(y)$
with a Gauss function in the study of 
$\langle (Q_{\Delta y})^n \rangle_{\rm c}$ up to the fourth order.
From the discussion in Sec.~\ref{sec:DME},
this result also suggests that the thermal blurring effect
can be described by the same manner as those developed in
Refs.~\cite{Kitazawa:2013bta,Kitazawa:2015ira}.

The result in Fig.~\ref{fig:Dy} is obtained for 
the fixed initial density $\rho_{\rm Y}(Y)$ without
event-by-event fluctuation.
When the event-by-event fluctuations of $\rho_{\rm Y}(Y)$ are included, 
the $\Delta y$ dependence of $R_n(\Delta y)$ is modified depending on 
parameters specifying the fluctuations of $\rho_{\rm Y}(Y)$.
Because this analysis is essentially the same as those 
addressed in Refs.~\cite{Kitazawa:2013bta,Kitazawa:2015ira}, 
in the present study we just refer to Figs.~2 and 3 in 
Ref.~\cite{Kitazawa:2013bta} and Figs.~4--8 in 
Ref.~\cite{Kitazawa:2015ira}, which show these results.
An important remark on these results is that 
with the inclusion of the event-by-event fluctuations of 
$\rho_{\rm Y}(Y)$, the thermal blurring effects on $R_3(\Delta y)$ 
and $R_4(\Delta y)$ can be enhanced significantly 
compared with the results in Fig.~\ref{fig:Dy}.

\subsection{Centrality dependence}
\label{sec:centrality}

\begin{figure}
\begin{center}
\includegraphics[width=.49\textwidth]{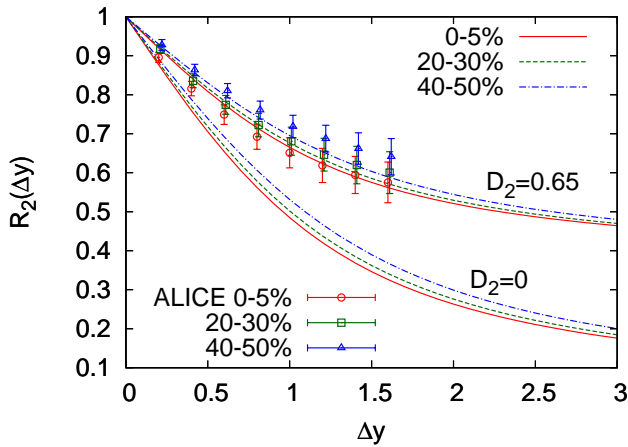}
\caption{
Centrality dependence of the second order cumulant of 
net-electric charge and the comparison with the experimental 
result by ALICE collaboration \cite{ALICE}.
}
\label{fig:ALICE}
\end{center}
\end{figure}

Next, we investigate the centrality dependence of 
the thermal blurring effect for the second order cumulant.
In the previous subsection we used parameters of the blastwave model, 
$T_{\rm kin}$ and $\beta$, for the most central collisions.
Because these parameters have centrality dependences 
\cite{Abelev:2013vea}, when one applies our results to 
non-central collisions the freezeout parameters have to be 
replaced with those corresponding to the centrality.
The experimental results suggest that $T_{\rm kin}$ becomes larger 
while $\beta$ becomes smaller from central to peripheral collisions
\cite{Abelev:2013vea}.
The analysis in Sec.~\ref{sec:n(y)} suggests that both these
dependences enhance the width of $n(y)$, and accordingly
the thermal blurring effect.

In this subsection, we include the event-by-event 
fluctuations of $\rho_{\rm Y}(Y)$ in the analysis
in order to compare the results with experimental data, 
as it is known 
that $\rho_{\rm Y}(Y)$ has such fluctuations in the early stage 
\cite{Asakawa:2000wh,Jeon:2000wg}.
We assume that the correlation function of $\rho_{\rm Y}(Y)$ 
has the form
\begin{align}
\langle \delta\rho_{\rm Y}(Y_1) \delta\rho_{\rm Y}(Y_2) \rangle_0
= D_2 \delta( Y_1-Y_2 ) \langle \rho_{\rm Y}(Y_1) \rangle_0.
\label{eq:<drdr>}
\end{align}
Note that Eq.~(\ref{eq:<drdr>}) is satisfied in the equilibrated
medium \cite{Asakawa:2015ybt}, and would be well justified even
near the QCD critical point \cite{Ling:2015yau}.
Here, $D_2$ is a quantity which is proportional to susceptibility
in the early stage \cite{Kitazawa:2015ira} and is related to 
D-measure $D$ \cite{Koch:2008ia} as $D_2=D/4$.

The results in the previous sections without fluctuations of $\rho_{\rm Y}(Y)$ 
is obtained with $D_2=0$, while in the hadron resonance gas model
one has $D_2\simeq1$ \cite{Koch:2008ia,Asakawa:2015ybt}.
By substituting Eq.~(\ref{eq:<drdr>}) into Eq.~(\ref{eq:Q2fluc})
and assuming uniform average density $\langle \rho_{\rm Y}(Y) \rangle_0=\rho_0$,
one obtains \cite{Shuryak:2000pd}
\begin{align}
\langle (Q_{\Delta y})^2 \rangle_{\rm c}
&= \rho_0 \int dY \left( \xi_2( p_{\Delta y}(Y) ) 
+ D_2 (p_{\Delta y}(Y))^2 \right)
\nonumber \\
&= \rho_0 \big[ 1 - (1-D_2) \int dY (p_{\Delta y}(Y))^2 \big].
\label{eq:Q_D2}
\end{align}

In Fig.~\ref{fig:ALICE}, we plot the $\Delta y$ dependence of 
$R_2(\Delta y)$ with blastwave parameters for centrality
bins $0-5\%$, $20-30\%$, and $40-50\%$ for ALICE experiment 
\cite{Abelev:2013vea} for $D_2=0$ and $0.35$.
In the figure, we also show the D-measure observed by 
ALICE collaboration with a translation $R_2(\Delta y)=D/4$
\cite{Koch:2008ia}.
Here, we emphasize that $D_2$ is the D-measure in the initial condition,
while $R_2(\Delta y)$ is the experimentally observed one with 
a rapidity window $\Delta y$, which takes a different value from $D_2$
owing to thermal blurring.
The figure shows that the results for $D_2=0.35$ agrees 
with the experimental data within the error for all centrality bins.
More accurate experimental data, however, is required to 
obtain a more quantitative conclusion.
It, however, is notable that the qualitative centrality 
dependence observed in Ref.~\cite{Abelev:2013vea}
is already well reproduced by thermal blurring
and centrality independent $D_2$.

\section{Blurring after diffusion}
\label{sec:D+B}

Up to now, we have estimated the magnitude of 
thermal blurring assuming that the particles are emitted 
from the hot medium at kinetic freezeout time.
In this argument, the distribution of $\rho_{\rm Y}(Y)$ in $Y$ space at 
{\it kinetic freezeout} is related to the experimentally-observed 
one after thermal blurring.
On the other hand, the experimentally-observed fluctuations are 
usually compared with thermal fluctuations in earlier stage, 
such as chemical freezeout time or much earlier, in the literature.
In this case, the modification of the event-by-event fluctuations 
in a rapidity window $\Delta Y$ before kinetic freezeout has to be 
taken into account besides the thermal blurring effect.
We emphasize that the coordinate-space rapidities $Y$ 
of individual particles and 
accordingly $\rho_{\rm Y}(Y)$ in each event are changing 
before the kinetic freezeout because of the nonzero velocity of 
individual particles along longitudinal direction 
\cite{Asakawa:2000wh,Jeon:2000wg,Shuryak:2000pd}.

If the motion of particles in $Y$ space before kinetic freezeout 
is well approximated by diffusion process,
a particle at $Y=Y_0$ at some early proper time $\tau=\tau_0$
is distributed at kinetic freezeout time $\tau=\tau_{\rm kin}$ 
in $Y$ space by a Gauss distribution 
\begin{align}
P_{\rm drift}(Y_0\to Y_{\rm kin}) 
\sim \exp\bigg( -\frac{ ( Y_{\rm kin} - Y_0 )^2 }{2d^2} \bigg),
\label{eq:P_drift}
\end{align}
with the diffusion distance $d$. 
Note that $d$ is related to the $\tau$ dependent 
diffusion coefficients $D(\tau)$ and $D_{\rm Y}(\tau)$ in cartesian and 
Bjorken coordinates, respectively, as
\cite{Sakaida:2014pya,Kitazawa:2015ira}
\begin{align}
d^2 = 2 \int_{\tau_0}^{\tau_{\rm kin}} d\tau' \frac{D(\tau')}{\tau'^2}
= 2 \int_{\tau_0}^{\tau_{\rm kin}} d\tau' D_{\rm Y}(\tau').
\end{align}

After the diffusion in $Y$ space, particles are observed 
at some rapidity $y$ through thermal blurring.
Then, the probability that a particle located at $Y=Y_0$ 
at $\tau=\tau_0$ is found at a rapidity $y$ in the final state
after thermal blurring, $P_{\rm D+B}(Y_0 \to y)$, is given by 
the convolution integral
\begin{align}
P_{\rm D+B}( Y_0 \to y ) 
= \int d Y_{\rm kin} P_{\rm drift}(Y_0\to Y_{\rm kin}) 
n(y - Y_{\rm kin}) .
\label{eq:D+B}
\end{align}

Generally, the probability Eq.~(\ref{eq:P_drift}) of the 
diffusion motion for different particles 
can be correlated because the scattering of particles 
can give rise to such a correlation.
We also note that the above argument does not take account of the 
possibility of pair creations and annihilations of particles.
When one considers the baryon number for sufficiently large 
$\sqrt{s_{NN}}$, however, the correlation should be well 
suppressed because baryons in the hadronic medium almost exclusively 
interact with pions \cite{KA}.
The chemical freezeout picture also suggests that the pair creations
and annihilations hardly occur after chemical freezeout time.
When these conditions are satisfied,
the total effect due to the diffusion in $Y$ space and thermal 
blurring can be described by simply replacing 
$n(y - Y)$ in Eq.~(\ref{eq:<Q^n>}) with $P_{\rm D+B}( Y_0 \to y )$
in Eq.~(\ref{eq:D+B}).
As discussed in Sec.~\ref{sec:num-dy}, the effect of 
non-Gaussianity of $n(y)$ is well suppressed.
By approximating $n(y)$ by a Gauss distribution,
$P_{\rm D+B}(Y_0\to Y_{\rm kin})$ given by the convolution of two 
Gauss ones also becomes Gaussian.
The total effect due to the diffusion and blurring then can be regarded 
as if it were from a single diffusion process, 
although the diffusion length, or the width of the Gauss distribution, 
includes both effects.
In this picture, the results in Figs.~\ref{fig:Dy} and \ref{fig:ALICE}
should be interpreted as the results with minimal diffusion lengths.
We also note that the hadronic decays \cite{Nahrgang:2014fza} 
give rise to diffusion of charges in rapidity space, 
and thus would be treated as a part of the diffusion and blurring 
to a good approximation.
Finally, we note that the same conclusion on thermal blurring is 
also applicable to the interpretation of the balance function 
and correlation along rapidity direction 
\cite{Pratt:2012dz,Gavin:2016hmv} measured experimentally.

\section{Summary}
\label{sec:summary}

In this study, we investigated the thermal blurring effect, 
i.e. the effect arising from the use of rapidity $y$ 
in substitution for the coordinate-space rapidity $Y$, 
on cumulants of conserved charges measured by the event-by-event 
analysis in relativistic heavy-ion collisions quantitatively.
Our analysis suggests that the thermal blurring affects 
fluctuation observables significantly at the maximal rapidity 
coverage of STAR detector, $\Delta y=1.0$, and not negligible 
even with $\Delta y=1.6$ the maximal rapidity coverage
of ALICE detector.
When one compares the event-by-event fluctuation observed in 
these experiments with theoretical results obtained on the basis of 
statistical mechanics, therefore, the correction arising from 
the thermal blurring effect should be taken into account seriously.

Although we assumed the Bjorken space-time evolution 
throughout this paper, at low energy collision this picture
does not hold any more. For lower energy collisions the effect of 
global charge conservation will also show up 
\cite{Bleicher:2000ek,Sakaida:2014pya}.
These effects have to be considered seriously in the interpretation
of experimentally-observed fluctuations at BES-II energy region
\cite{BES-II}.
Throughout this study we also assumed that the transverse momentum 
is integrated out. In real experiments, however, the particles are
observed in a finite transverse momentum acceptance. 
The understanding of the effect of the momentum cut \cite{Karsch:2015zna} 
on fluctuations is another important issue.
The comparison and estimate of the thermal blurring effects
based on the dynamical models \cite{Herold:2016uvv} are also 
interesting subjects for future study.

The authros thank the discussion with M.~Sakaida.
This work was supported in part by 
JSPS KAKENHI Grant Numbers 25800148, and 26400272.

\end{document}